\documentclass[journal=jacsat,manuscript=article]{achemso}
\usepackage[version=3]{mhchem}

\usepackage{braket}
\usepackage{hyperref} 
\usepackage{xcolor}
\usepackage{textcomp}
\usepackage{gensymb}

\author{Arsineh Apelian}
\affiliation{Materials Department, University of California, Santa Barbara, CA 93106-9510, U.S.A.}
\author{Annabelle Canestraight}
\affiliation{Department of Chemical Engineering, University of California, Santa Barbara, CA 93106-9510, U.S.A.}
\author{Songyuan Liu}
\affiliation{Department of Physics, University of California, Santa Barbara, CA 93106-9510, U.S.A.}
\author{Vojt\v{e}ch Vl\v{c}ek}
\alsoaffiliation{Materials Department, University of California, Santa Barbara, CA 93106-9510, U.S.A.}
\affiliation{Department of Chemistry and Biochemistry, University of California, Santa Barbara, CA 93106-9510, U.S.A.}
\email{vlcek@ucsb.edu}

\title[Quasiparticles in moir\'e systems]
  {Delocalization of Quasiparticle Moir\'e States in Twisted Bilayer hBN}
  
\abbreviations{IR,NMR,UV}

\begin{document}

%TC:ignore
\begin{abstract} Twisted bilayers host many emergent phenomena in which the electronic excitations (quasiparticles - QPs) are closely intertwined with the local stacking order. By inspecting twisted hexagonal boron nitride (t-hBN), we show that non-local long-range interactions in large twisted systems cannot be reliably described by the local (high-symmetry) stacking and that the band gap variation (typically associated with the moir\'e excitonic potential) shows multiple minima with variable depth depending on the twist angle. We investigate twist angles of 2.45$^\circ$, 2.88$^\circ$, 3.48$^\circ$ and 5.09$^\circ$ using the GW approximation together with stochastic compression to analyze the QP state interactions. We find that band-edge QP hybridization is suppressed for intermediate angles which exhibit two distinct local minima in the moir\'e potential (at AA region and saddle point (SP)) which become degenerate for the largest system (2.45$^\circ$).

\end{abstract}
%TC:endignore

Twisted van der Waals structures have attracted considerable attention in the past decade \cite{Schaibley2016, Pan2018, Yankowitz2019, Zheng2020, Leconte_2020}. These systems, composed of nominally weak monolayers, exhibit strongly correlated behavior\cite{PhysRevB.85.195404, Yankowitz2012} and harbor strongly localized excitations (quasiparticles - QPs), i.e. charged electrons, holes, or excitons, subject to a periodic potential landscape whose energy and spatial distribution are determined by the twist angle. It is widely recognized that the moir\'{e} superlattice can modulate the electronic band structure \cite{doi:10.1021/nl501077m} and exhibit emergent properties such as unconventional superconductivity \cite{supercon} and insulating behaviour driven by correlations \cite{dirac, kim, Dean2013}. Similar new quantum phases are predicted for twisted bilayers of hexagonal boron nitride (t-hBN) \cite{doi:10.1021/acs.nanolett.9b00986, PhysRevLett.124.086401, PhysRevB.108.075109}, transition-metal dicalcogenides (TMD) \cite{Devakul2021, Wang2020, https://doi.org/10.1002/sstr.202000153, PhysRevLett.122.086402, PhysRevB.102.241106, Su2022, Chan2024}, or phosphorene (t-BP) \cite{Brooks_2020, PhysRevB.105.235421, Zhao2021, UdDin2022, PhysRevB.103.L041407}.

Guiding the development of moir\'{e} platforms necessitates quantitative theoretical understanding of their properties. Commonly, they are derived from the local arrangements that are conveniently represented by the parent (``high-symmetry'') structures. The QP's in such bilayers are subject to a local potential landscape formed by the distinct stacking orders in the 2D lattice. This is invoked in the description of effective excitonic potentials based on the local band gaps (i.e., local differences between QP energies of an electron and a hole) \cite{Mac1, Mac2, shedding, PhysRevB.102.201115}. Experiments show that by optically exciting the localized moir\'{e} states can generate strong interactions between electrons and holes, making them act as deep potential wells for excitons \cite{moody, Seyler2019, Jin2019, Tartakovskii2020}. Here, the exciton is typically described as a QP moving through a slowly-varying potential energy landscape which is often characterized by the variation of the local band gaps because the band-edge energy changes with the local geometry of the moir\'{e} superlattice \cite{Li2021, Mac1, Mac2}. 

The extent of the non-local long-range interactions and effect on the individual QPs in the moir\'{e} potential is debated, given that localization is further aided through lattice distortions, enhancing the ``trapping'' of electronic states \cite{Naik2018}. First-principles calculations of the twisted systems require large supercells with thousands of atoms, and they are thus typically limited to mean-field methods. Recent efforts in many-body perturbation theory (MBPT) focus on reconstructing the correlations from calculations of the parent systems \cite{doi:10.1126/science.add9294, Naik2018, Naik2022, Li2021} and rely on mean-field orbitals. To our knowledge, the only QP (Dyson) orbital reconstruction was so far achieved with stochastic methods for the correlated subspace in twisted bilayer graphene \cite{Romanova2022}. Moreover, the role of nonlocal correlations on the spatial extent of Dyson orbitals and their hybridization is unknown.  

We overcome these limitations and investigate the QP landscapes by employing the $GW$ approximation in the stochastic MBPT formulation \cite{vlcek, vlcek2017stochastic}. To study the moir\'{e} states, we extend the technique to determine the Dyson orbitals from a stochastic compression of the orbital basis \cite{annabelle}. We explore the QP energy variation within the moir\'{e} lattice and the hybridization of the orbitals for t-hBN whose electronic structure is well described by $GW$ \cite{PhysRevB.108.165108} where the non-local correlation effects are considered and thus include weak (van der Waals) interlayer interactions. Further, calculations reveal the formation of flat bands and strong spatial localization of electrons at small twist angles \cite{doi:10.1021/acs.nanolett.9b00986}. Experiments also confirm excitonic states in t-hBN \cite{roux2024exciton}. 

Here, we consider four small twist angles: $2.45^{\circ}$, $2.88^{\circ}$, $3.48^{\circ}$ and $5.09^{\circ}$, with structures containing up to $\approx$ 18,000 electrons. Our results reveal how localized moir\'{e} states hybridize due to a nontrivial interplay between local structure and non-local correlations. We compare the band gaps obtained independently for the individual high-symmetry structures with those performed on the entire t-hBN system, evaluated using the localized orbitals in various regions of the bilayer. The latter shows significantly reduced spatial variations, with the differences between adjacent low-gap regions (valleys) decreased by almost an order of magnitude. For most twist angles, the nonlocal interactions lead to band-edge QP states mixing (hybridization), which is suppressed for intermediate angles. This behavior is mostly determined by the strength of interactions and localization in high-symmetry regions of the moir\'{e} cell. Most importantly, the calculations for the twisted systems reveal that the local excitonic potential is significantly decreased in the SP stacking region, leading to a characteristic ``double well''.

T-hBN, a wide band gap semiconductor \cite{Cassabois2016}, has a honeycomb-like structure and periodicity on the nm scale. Unlike other systems (e.g., TMD's \cite{PhysRevLett.108.196802, Mak2012, Zeng2012, Cao2012, Jones2013}), it exhibits negligible spin-orbit coupling (SOC) and serves as a simple parent material to investigate at what extent can one approximate the moir\'{e} potential according to only the local environments and when it becomes necessary to consider fully twisted structures. 

\begin{figure*}[ht]
    \centering
    \includegraphics[width=1.0 \linewidth]{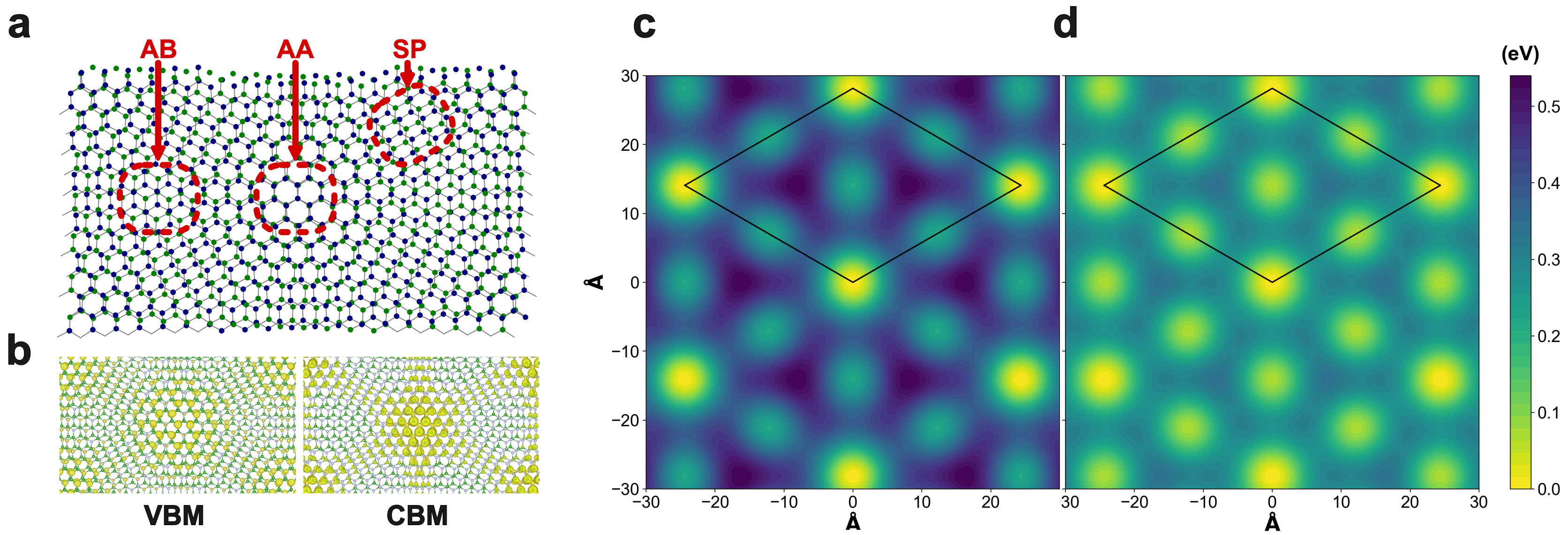}
    \caption{\textbf{a)} Structure of t-hBN with a twist angle of $5.09^{\circ}$ showing the high-symmetry regions in dashed red boxes (AA, AB, SP) \textbf{b)} LDOS for the states near the valence band maxima (VBM) and conduction band minima (CBM) showed as a density isosurface in yellow. \textbf{c)} Local band gap variation (computed with  s-$G_{0}W_{0}$ for $\theta = 5.09^{\circ}$) translated to a moir\'{e} potential heatmap constructed from the local environments based solely on the high-symmetry stacking regions. The deep minima correspond to the AA region (set to zero on  the heat map scale); the moir\'e unit cell is shown for comparison. \textbf{d)} Moir\'e potential constructed from the LDOS projection computed for the entire twisted system shows much weaker local variations compared to the local approximation in \textbf{c)}.
 }
    \label{fig:Fig1}
\end{figure*}

To disentangle the role of non-local correlations (stemming from density-density fluctuations) and the local structural arrangements, we first consider the electronic structure of the full moir\'{e} system and compare it to results for untwisted structures based on the high-symmetry stackings. We deliberately construct non-corrugated (flat) bilayers in which the QP localization is driven by the potential variations due to the atomic position mismatch in the two layers. Here, we investigate the electronic contributions to QP localization, as opposed to distortion-assisted trapping of electronic states, which will further contribute to the localization of electronic states.     

Three high-symmetry points (AA, AB, SP) are identified as depicted in Fig.\ref{fig:Fig1} (a) and twist angles of $2.45^{\circ}$, $2.88^{\circ}$, $3.48^{\circ}$ and $5.09^{\circ}$ are considered. First, we study how the electronic states and band gaps are impacted by the interactions in twisted systems and explore the spatial distribution of the charge density in the band-edge states, i.e., the local density of states (LDOS), constructed from a set of nearly degenerate electronic states which connect the relevant high-symmetry points (all those are within $\sim$ 140 meV for valence and  $\sim$100 meV for conduction states). We separate the states corresponding to the individual stacking regions by their spatial distributions. The degenerate edge states at the valence band show AA localization shown in Fig. \ref{fig:Fig1}(b)(left), which is in agreement with previous works \cite{AAstack}. Energetically more distant valence states exhibit localization in the immediate neighborhood of AA, and deeper states span AB and SP regions (see Fig. S3 in SI). For the conduction band, the degenerate band-edge states (up to $\sim$ 100 meV) exhibit hybridization of two different stacking regions where electrons are localized in both the AA and SP regions shown in Fig. \ref{fig:Fig1}(b)(right). Already at this point, the calculation for the entire t-hBN exhibits states delocalized between two high-symmetry points, which is missed when only local stacking orders are considered.

Despite the hybridization, we define the local band gaps by associating it with the transitions between the electronic states that are localized in the corresponding high-symmetry regions (i.e., relating  the LDOS and QP energies). We compare these results with those obtained from the calculations for untwisted parent structures. We first employ the common ``one-shot'' perturbative correction, stochastic $G_{0}W_{0}$ \cite{vlcek, vlcek2017stochastic}, to the DFT eigenvalues. The computed $G_0W_0$ band gaps for t-hBN opens up to roughly the experimental value of approximately 6 eV \cite{Watanabe2004, Paleari_2018} (depending on the particular twist angle or stacking). Full results (including DFT) for each region can be found in Table S.1 in SI. As discussed later, the one-shot corrections are sufficient to address the energies of the individual states, though they rely on the spatial distribution of the underlying KS orbitals (we address the QP state hybridization later).

From these results, we construct the effective excitonic moir\'{e} potential energy $\Delta(\textbf{r})$ which is commonly applied to describe the twisted bilayers\cite{Mac1}. Here, the exciton can be described as a QP (an entangled quasi-electron and quasi-hole pair) in a potential landscape dominated by the variation of the local band gaps (assuming that the exciton binding energy is roughly constant across the system, following \cite{Mac1}). In contrast to the typical construction of $\Delta(\mathbf{r})$, we employ a second-order harmonic expansion to account for the SP regions in the moir\'{e}, resulting in an additional local extrema:
\begin{equation}
    \Delta(\textbf{r}) = V_0 +  \sum_{j=1,2,3} [V_{1}\cos(\textbf{b}_{j}\cdot\textbf{r}) + V_{2}\cos(2\textbf{b}_{j}\cdot\textbf{r})].\label{eq:moire_potential}
\end{equation}
Here, $\left| \textbf{b}_{j} \right |= \frac{4\pi}{\sqrt{3}a_{M}}$ scales the moir\'{e} 
reciprocal lattice vectors where $a_{M}$ is the moir\'{e} period. The constant $V_0$ term determines the average gap of the twisted system while the local fluctuations are captured by $V_1$ and $V_2$. Note that if only the AA and AB regions are considered (no second local minimum is present), $\Delta(\textbf{r})$ is approximated by the lowest-order harmonic expansion\cite{Mac1}.

The comparison of the gap variations in the  twisted structures with those obtained solely from the local environments differ significantly. The latter exaggerates the variation and the barrier heights between the distinct stacking points, as shown in Figs. \ref{fig:Fig1}(c) and (d) for the $\theta = 5.09^{\circ}$ system. When considering only the high-symmetry regions, the depth of the moir\'{e} valleys are overestimated by up to $0.28$ and 0.13~eV and for the AA and SP regions and $\Delta(\mathbf{r})$ computed for the entire system is rather uniform with shallow valleys.

\begin{figure*}[ht]
    \centering
    \includegraphics[width=0.99 \linewidth]{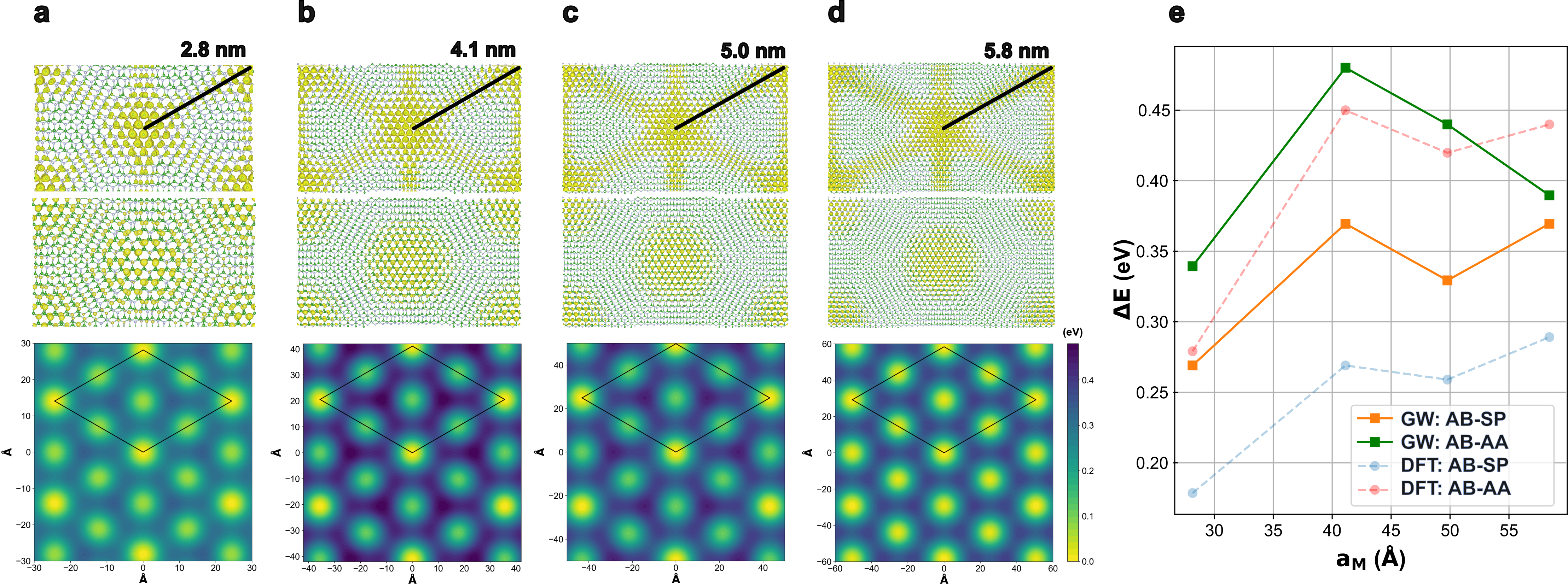}
    \caption{Plotted LDOS of the CBM (top) and VBM (bottom), and the corresponding moir\'e potentials for twist angles of \textbf{a)} 5.09$^\circ$ \textbf{b)} 3.48$^\circ$ \textbf{c)} 2.88$^\circ$ \textbf{d)} 2.45$^\circ$. The zero of the moir\'e potential is set at the minimal local gap in the AA stacking region (yellow on the heat map). The moir\'e unit cell is shown for each plot by black lines with vertices in the AA region. At 3.48$^\circ$ the system shows weak secondary minima in the SP region, while it is significantly more pronounced for 2.45$^\circ$. \textbf{e)} The distance between the local minima (AA or SP) and the maximum (AB) of the moir\'e potential as a function of the moir\'e lattice constant $a_M$. The AB-AA curve shows a maximum for intermediate twist angles; for large $a_M$ the AA and SP minima become practically degenerate, leading to two pronounced minima (visible in \textbf{d)}). }
    \label{fig:Fig2}
\end{figure*}

Next, we investigate how the moir\'e potential and mutual interactions of localized states change with the twist angle and consider only the full t-hBN system. We find that decreasing the mutual twist angle does not reduce the apparent delocalization of electronic states, despite the larger moir\'e period and increased distance between high-symmetry regions. This is somewhat surprising as one might expect that the localized states will get more separated and effectively isolated. ``Hybridization'' of the high-symmetry regions, particularly in the conduction states, is clearly observed whether there is a twist angle of 5.09$^{\circ}$ or a twist angle of 2.45$^{\circ}$ (see Figs. \ref{fig:Fig2}(a)-(d) (top)). Computed $\Delta(\mathbf{r})$ for all twist angles is shown in Figs. \ref{fig:Fig2}(a)-(d) (bottom)). The SP valley is deeper for small twist angles (most pronounced for $\theta = 2.45^{\circ}$) and practically indistinguishable (within the stochastic error) from the results for AA. In essence, the small-angle band gap landscape will show two sets of practically identical minima localized in distinct stacking regions. Depending on the twist angle, the interlayer excitons thus can be confined similarly by both AA and SP regions. 

Another surprising observation is that the barrier between SP and AB regions is highly non-monotonic as a function of the moir\'e lattice constant $a_M$. The energy differences between AA and AB and similarly between SP and AB are shown in Fig. \ref{fig:Fig2}(e) with maximum for intermediate angles for AA-AB. The bilayer with $\theta = 3.48^{\circ}$ has the greatest variation between the different regions; it has the deepest potential in the AA region while a relatively shallow well at SP. Notably, this is associated with increased localization of electronic states in the AA region which cannot be simply inferred from the visual inspection of the LDOS, and it is discussed below. As these effects are to a large extent driven by the nonlocal self-energy, the evaluation of $\Delta (\mathbf{r})$ with the twist angle is missed by DFT. For $\theta=2.45^\circ$, DFT results show a much larger difference of the $\Delta (\mathbf{r})$ depth, i.e., the gaps at AA and SP region differ by $\sim 0.15$~eV as opposed to $\sim 0.02\pm 0.04$~eV with s-$G_0W_0$. Further, DFT does not show a non-monotonic dependence of the valley barrier heights and both AA and SP minima are increasing towards small twist angles with almost constant energy offset (Fig.~\ref{fig:Fig2} (e)).

To explore the coupling between different moir\'{e} states and the differences between KS and QP orbitals, we next investigate the effective QP Hamiltonian represented in the KS basis. We will focus on the role of off-diagonal self-energy terms where $\Sigma_{jk}(\omega) = \braket{\phi_{j}|\Sigma(\omega)|{\phi_{k}}}$ which mixes the KS orbitals. For the fixed-point solutions eigenstates are the many-body Dyson orbitals that incorporate the nonlocal exchange and dynamical correlations. Given the large number of QP states (up to $8,752$ occupied states) and aim to facilitate the comparison among different bilayers with a different number of total electrons, we employ a modified stochastic orbital compression technique\cite{annabelle}. Here, multiple orbitals are represented simultaneously by random vectors in the single-particle Hilbert space (obtained from the diagonalization of the KS Hamiltonian), selected such that they sample predetermined subspaces. This drastically reduces the computational cost and provides accurate estimates of QP energies when hybridization is involved \cite{annabelle}.

\begin{figure*}[ht]
    \centering
    \includegraphics[width=0.9 \linewidth]{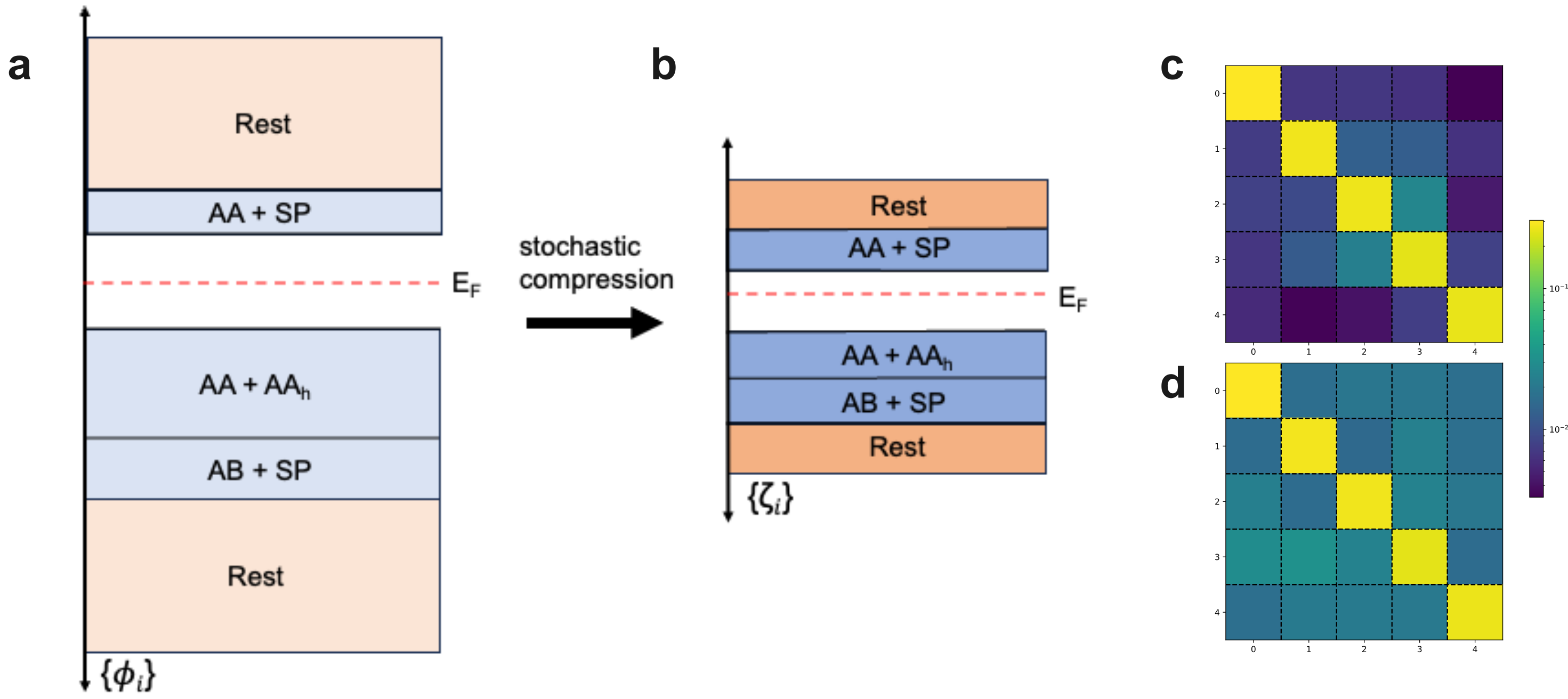}
    \caption{\textbf{a)} Illustration of the single-particle KS states distribution in the subspaces corresponding to the various high-symmetry stacking regions: AA, AB, SP. The AA$_{\rm h}$ region corresponds to the hybridized localization in the immediate neighborhood of AA.  \textbf{b)} The compression of the states into a reduced dimensional representation with only five regions, among which the coupling is analyzed. \textbf{c)} Visualization of the full QP Hamiltonian matrix in the KS basis for the compressed states for the $3.48^{\circ}$ twist angle that is diagonally dominant and hence localized as in the KS picture. The rows from top to bottom and left to right represent the states of increasing energy (i.e., the top row corresponds to the block of the rest occupied states. In contrast, \textbf{d)} shows the QP Hamiltonian for $2.45^{\circ}$ for which the QP states further hybridize more strongly .} 
    \label{fig:Fig3}
\end{figure*}

We choose a subspace of states that are identified through their localization in LDOS (Fig.~\ref{fig:Fig3}(a)) and sample them by random vectors $\{\zeta\}$. We divide the system into a set of valence states that correspond to: AA localization, states in the immediate neighborhood of AA (AA$_{\rm h}$), and AB and SP hybridization as well as a set of conduction states that correspond to the AA and SP hybridization. Instead of treating all of them explicitly, we construct stochastic orbitals: $\ket{\zeta} = \sum_{j}^{N_{\zeta}}c_{j} \ket{\phi_{j}}$ where $N_{\zeta}$ is the number of random vectors that we choose and which are mutually orthogonalized. We employ $N_\zeta = 14$ states (with at least 4 random vectors sampling each subspace) to sample the off-diagonal elements of the self-energy corrections. $N_\zeta$ was selected such that the expectation values of the underlying KS Hamiltonian reproduced the deterministic values with a statistical error $<0.02\%$. This estimate is based on the comparison of the equivalent (known) KS Hamiltonian entries $H^{KS}_{jk}$  and selected self-energy entries $\Sigma_{jk}$. This compression uncovers the degree of hybridization between nearly degenerate states localized in the same stacking regions (see also ref. [\cite{annabelle}]). For completeness, we also include sampling of the states outside of the subspace of interest within the energy distance of 1.1~eV. These ``rest states'' are sampled by $N_\zeta^{\rm rest} = 4$ random vectors for $\theta = 3.48^{\circ}$ and $N_\zeta^{\rm rest} = 8$ random vectors for $\theta = 2.45^{\circ}$. Our results are practically independent of $N_\zeta^{\rm rest}$, i.e., the high-symmetry regions are not further hybridizing with the energetically distant states. This random vector basis is used to evaluate the QP Hamiltonian $H^{QP}_{jk}(\omega)$ where $j$ and $k$ represent state indices and the off-diagonal terms ($j\neq k$) in the random basis thus represent the coupling between the individual subspaces.

A full-frequency treatment is executed when computing the off-diagonal self-energies. The QP Hamiltonian matrix is evaluated and diagonalized at a dense frequency mesh (with the spacing of 0.01 atomic units), and the fixed-point solutions $(\omega_{i}=\epsilon_{i})$ are found with linear interpolation on the $\omega$ grid. In all cases, the QP energy is well defined by a single fixed-point solution (pole of the GF). As the Dyson orbitals are generally obtained by diagonalization at distinct frequencies, they are not orthogonal \cite{martin_book}. For visualization and analysis of the couplings, we averaged each component from each matrix across all 22 Hamiltonians satisfying the QP fixed-point solutions, resulting in a single $22 \times 22$ self-energy matrix for $\theta = 2.45^{\circ}$, for instance. Lastly, due to the several degenerate states in the twisted bilayers, we perform an additional compression by averaging elements from each set of states, resulting in a (5$\times$5) compressed QP Hamiltonian matrix as shown in Fig. \ref{fig:Fig3}(c) for $\theta = 3.48^{\circ}$ and (d) for $\theta = 2.45^{\circ}$, corresponding to the systems with the largest and lowest energy differences between the AA and SP $\Delta(\mathbf{r})$ potential minima. Note that the basis compression for the diagonalization does not truncate the self-energy, which contains information on all the single-particle states through Green's function and the screened Coulomb interaction. Here, each block represents an averaged subspace of states containing: rest (occupied) states, AB+SP, AA+AA$_{\rm h}$, AA+SP, and rest (unoccupied) states. 

We find that the QP energies of $H^{QP}$, are mostly diagonal, and the off-diagonal eigenvalues differ only negligibly from those computed within the diagonal approximation; for instance, the maximum deviations are $1.33\%$ difference for $\theta = 3.48 ^{\circ}$ and $6.65\%$ for $\theta = 2.45^{\circ}$ for the localized states of interest. Still, we find a stark difference in the off-diagonal terms (determining the mixing of the KS states). Most importantly, for $\theta = 3.48^{\circ}$, the QP Hamiltonian shows weaker mixing of the underlying KS states (despite some is observed across the gap in for the AA localized states indicated by lighter color of the off-diagonal block of $H^{QP}$ in the third and fourth row/column in Fig. \ref{fig:Fig3}(c)). These results confirm the findings for the QP energies presented above and also further strengthen the observation that systems with intermediate twist angles (here $\theta = 3.48^{\circ}$) harbor less hybridized, i.e., less delocalized, states compared to small or large twists.

%By studying the near-gap electronic states (and band gap variations) in t-hBN, we uncover how the electronic structure of large-scale twisted structures depends on the underlying non-local interactions. The local gap variation (associated with the excitonic potential) cannot be simply generated from considering only local environments. The latter leads to a gross exaggeration of the features and spuriously suggests a strong confinement of excitons only in one region. 

In conclusion, by studying the near-gap electronic states and band gap variations in t-hBN, we uncovered how the electronic structure of large-scale twisted structures depends on the underlying non-local interactions. The local gap variation (associated with the excitonic potential) cannot be simply generated from considering only local environments. The latter leads to a gross exaggeration of the features and spuriously suggests a strong confinement of excitons only in one region. There also exists a nontrivial interplay of correlations between the monolayers, which significantly reduces the energy barriers between the valleys of the excitonic potential. For large twist angles, the small moir\'{e} unit cells lead to strong hybridization of the localized states and rather uniform band gap variation. At small twist angles, the correlation effects weaken, but the energy barriers are much smaller, again leading to delocalization of electronic states across the high-symmetry regions. Surprisingly, there is an intermediate twist angle regime at $\theta = 3.48^{\circ}$ at which the energy separation barriers between the individual localized states are large. We surmise that for such angles, the QPs localization may not need to be assisted by phonons or lattice distortions. Further, we find that the moir\'e potential hosts two minima in distinct stacking regions (which become practically degenerate for small twist angles) that can host two types of moir\'e excitons. Besides addressing the QP properties discussed in this work, the methodology allows the development of accurate \textit{ab-initio} model Hamiltonians to describe electronic, excitonic\cite{Mac1}, magnetic properties \cite{PhysRevLett.132.076503}, and other rich physics related to the strong correlations \cite{lu2022low, bennett2023twisted, PhysRevB.107.235131}. These models have so far heavily relied on the analysis of first-principles calculations based on solely the local environments. We believe this is an important stepping stone for reevaluating how we treat moir\'{e} superlattices, allowing for a more direct comparison to experiments.

%To study the hybridization between different moir\'{e} states (systems as large as $\sim$18,000 electrons), we expanded our stochastic theory to address the role of off-diagonal self-energies. Besides addressing the QP properties discussed in this work, the methodology allows the development of accurate \textit{ab-initio} model Hamiltonians to describe electronic, excitonic\cite{Mac1}, magnetic properties \cite{PhysRevLett.132.076503}, and other rich physics related to the strong correlations \cite{lu2022low, bennett2023twisted, PhysRevB.107.235131}. These models have so far heavily relied on the analysis of first-principles calculations based on solely the local environments. We believe this is an important stepping stone for reevaluating how we treat moir\'{e} superlattices, allowing for a more direct comparison to experiments.

\begin{suppinfo}
Computational details, stochastic many-body theory, band gap data, moir\'{e} potential fit, LDOS plots, self-energy Hamiltonian.

\end{suppinfo}

\begin{acknowledgement}
The authors thank Barak Hirshberg and Netanel Bachar Schwartz at Tel Aviv University for their fruitful discussions. The methodological development, selected simulations for the large-scale systems, and the analysis were supported by the National Science Foundation (NSF) CAREER award through grant No. DMR-1945098. We acknowledge funding from the UC Office of the President within the Multicampus Research Programs and Initiatives (M23PR5931) supporting some of the calculations performed in this work. The work on model moir\'e potentials was supported by the United States - Israel Binational Science Foundation (contract \#2020083). Use was made of computational facilities purchased with funds from the NSF (CNS-1725797) and administered by the Center for Scientific Computing (CSC). The CSC is supported by the California NanoSystems Institute and the Materials Research Science and Engineering Center (MRSEC; NSF DMR 2308708) at UC Santa Barbara. A.A. was supported by the NSF Graduate Research Fellowship under Grant No. (2139319).

\end{acknowledgement}

\bibliography{bib}

\end{document}